\documentclass[a4paper,11pt]{article}

\usepackage{amsmath,amssymb,bbm}
\usepackage{cite}
\usepackage[dvipdfm]{graphicx}

\makeatletter

 \@addtoreset{equation}{section}
\makeatother

\newcounter{Enumerate}

\DeclareFontFamily{U}{rsf}{}
\DeclareFontShape{U}{rsf}{m}{n}{
  <5> <6> rsfs5 <7> <8> <9> rsfs7 <10-> rsfs10}{}
\DeclareMathAlphabet\Scr{U}{rsf}{m}{n}

\usepackage[mathscr]{eucal}

\newcommand{\del}{\partial}
\newcommand{\half}{\frac{1}{2}}
\newcommand{\mfk}{\mathfrak}
\newcommand{\LS}{\ \ \ \ \ \ \ \ \ \ }
\newcommand{\ls}{\ \ \ \ \ }
\newcommand{\wt}{\widetilde}

\newcommand{\ol}{\overline}

\newcommand{\bsubeq}{\begin{subequations}}
\newcommand{\esubeq}{\end{subequations}}
\newcommand{\eps}{\epsilon}
\newcommand{\nn}{\nonumber}

\newcommand{\N}{\mathcal{N}}

\renewcommand{\d}{{\rm d}}
\newcommand{\e}{{\rm e}}
\renewcommand{\i}{{\rm i}}
\renewcommand{\l}{\ell}
\renewcommand{\t}{\mfk{t}}

\newcommand{\slb}{\scalebox}

\renewcommand{\Im}{{\rm Im}}
\renewcommand{\Re}{{\rm Re}}

\begin{document}
\allowdisplaybreaks{

\thispagestyle{empty}


\begin{flushright}
KEK-TH-1526 
\end{flushright}

\vspace{30mm}

\begin{center}
\slb{2}{On Static Charged Black Holes}

\vspace{5mm}

\slb{2}{in Type IIA on a Nearly-K\"{a}hler Coset}

\vspace{15mm}

\slb{1.2}{Tetsuji {\sc Kimura}} 

\vspace{2mm}

{\sl
KEK Theory Center,
Institute of Particle and Nuclear Studies, \\
High Energy Accelerator Research Organization \\
Tsukuba, Ibaraki 305-0801, Japan}

\vspace{1mm}

\slb{0.9}{\tt tetsuji\;\_at\_\;post.kek.jp}

\end{center}

\vspace{10mm}


\begin{abstract}
We study static, spherically symmetric black hole solutions in four-dimensional $\N=2$ gauged supergravity with one vector multiplet and one hyper-tensor multiplet.
This is derived from massive type IIA theory compactified on the nearly-K\"{a}hler coset space $G_2/SU(3)$.
It is well-known that the Romans' mass parameter yields the St\"{u}ckelberg-type deformation of the gauge field strengths in the four-dimensional system.
This deformation requires that all the (covariant) derivatives of the scalar fields must vanish and the two-form field is closed.
It turns out that charged solutions are forbidden. 
This implies that only AdS vacua or Schwarzschild-AdS black holes are allowed as the static, spherically symmetric solutions.

\end{abstract}

\newpage

\section{Introduction}
\label{sect-introduction}

Searching asymptotically Anti-de Sitter (AdS) black hole solutions in four-dimensional $\N=2$ gauged supergravity \cite{Andrianopoli:1996cm} has been developed for a long period.
This is quite interesting because the value of the cosmological constant in gauged supergravity is non-trivial; i.e.,
the cosmological constant is given by the Fayet-Iliopoulos parameters or the expectation value of the scalar potential.
In the context of AdS/CFT correspondence, the AdS black hole configuration provides many significant features of condensed matter physics.

In pure supergravity without any matter fields, 
supersymmetric (non)-rotating AdS black holes with (un)usual topology were investigated by \cite{Caldarelli:1998hg}. 
Soon after that, 
gauged supergravity with vector multiplets was applied to search static AdS black holes with naked singularity \cite{Sabra:1999ux, Chamseddine:2000bk, Bellucci:2008cb}.
Recently, supersymmetric static AdS black holes with regular horizon were found \cite{Cacciatori:2009iz, Hristov:2010ri, Dall'Agata:2010gj}.

It is also interesting to find an AdS black hole solution 
of gauged supergravity in the presence of hypermultiplets \cite{TK1108}.
There are two features.
One is that the existence of hypermultiplets excludes the Fayet-Iliopoulos parameters, which support non-vanishing cosmological constant in the gauged supergravity only with vector multiplets. 
The other is that
some scalar fields in hypermultiplets appears as two-form fields in the system. 
Caused by the gauging, they cannot be dualized back to the original scalar fields in the hypermultiplets.
The multiplet containing two-form field(s) is referred to as the hyper-tensor multiplet.
In addition, these two-form fields deform the gauge field strengths by the St\"{u}ckelberg-type coupling \cite{Dall'Agata:2003yr, D'Auria:2004yi}.
In order to control the feature of the gauged supergravity with multiplets of the hyper-sector as well as vector multiplets, 
we study AdS black holes in the framework of the string theory compactification scenarios.

Exploiting four-dimensional $\N=2$ gauged supergravity is of importance
because this system appears as the low energy effective theory of type II string theory via flux compactifications \cite{Grana:2005jc, D'Auria:2007ay, Cassani:2008rb, Cassani:2009na}.
The deformation parameters in gauging of the four-dimensional supergravity 
are provided by the NSNS- and the RR-flux charge parameters on the internal space.
Indeed the St\"{u}ckelberg-type deformation is realized by the RR flux charges \cite{D'Auria:2007ay, Cassani:2008rb, Cassani:2009na}.
In particular, the Romans' mass parameter yields AdS vacua \cite{Lust:2004ig}.
In this work we focus on the coset spaces $G_2/SU(3)$ \cite{KashaniPoor:2007tr, Cassani:2009ck}.
This is one of the nearly-K\"{a}hler manifold with torsion and the $SU(3)$-structure.
This is useful because 
the moduli space of this coset space has been studied very well.
In addition, this coset space provides the simplest matter contents in the four-dimensional system; one vector multiplet and one hyper-tensor multiplet with the St\"{u}ckelberg-type deformation.

We are now ready to search black holes of Reissner-Nordstr\"{o}m-AdS type
in four-dimensional $\N=2$ gauged supergravity with one vector multiplet and one hyper-tensor multiplet given by the flux compactification on $G_2/SU(3)$.
Previously the author studied it under the (covariantly) constant condition \cite{TK1108}.
This condition prohibits the existence of charged solutions.
In this paper, we assume that the configuration is static and spherically symmetric.
Surprisingly the static condition eventually derives the (covariantly) constant condition \cite{TK1108}.
This implies that the static condition itself forbids Reissner-Nordstr\"{o}m-AdS black hole solutions of the gauged supergravity in the presence of the two-form field with the St\"{u}ckelberg-type deformation \cite{Dall'Agata:2003yr, D'Auria:2004yi, D'Auria:2007ay, Cassani:2008rb, Cassani:2009na}.

The organization of this paper is as follows.
In section \ref{sect-profile}
we briefly exhibit the feature of the coset space $G_2/SU(3)$.
Next we write down the equations of motion for the gauge fields, the two-form field and the gravitational field which play a central role in the main analysis.
Third we restrict the system to the static configuration. 
Here we introduce two functions in order to describe the static field strengths.
In section \ref{sect-analysis}
we show that all the fields are (covariantly) constant in the static configuration:
The equations of motion for the gauge fields intertwine the two-form field with the two functions.
The equation for the two-form field provides the differential equations among the two functions and the scalar fields.
Finally, the Einstein equation reveals that all the scalar fields are (covariantly) constant because each term quadratic in the (covariant) derivatives of the scalar fields is positive semi-definite.
We also find that the two-form field is closed.
The (covariantly) constant condition only allows neutral solutions such as AdS vacua or Schwarzschild-AdS black holes.
In section \ref{sect-conclusion} is devoted to conclusion.


\section{Gauged supergravity with two-form field}
\label{sect-profile}

The deformation parameters in gauging of the four-dimensional supergravity 
are provided by the NSNS-flux charge parameters $\{ e_{\Lambda I} , e_{\Lambda}{}^I, m^{\Lambda}{}_I , m^{\Lambda I} \}$ and the RR-flux charge parameters $\{ e_{\text{R}\Lambda}, m_{\text{R}}^{\Lambda} \}$ on the internal space 
\cite{D'Auria:2007ay, Cassani:2008rb, Cassani:2009na}. 
The ranges of the labels $\Lambda$ and $I$ are 
$\Lambda = 0,1,\ldots,n_{\text{V}}$ and 
$I = 0,1,\ldots,n_{\text{H}}$, 
where $n_{\text{V}}$ denotes the number of the vector multiplets,
whilst $n_{\text{H}}$ indicates the number of multiplets in the hyper-sector, respectively.
The Romans' mass parameter is involved as $m_{\text{R}}^0$ in the above flux charge parameters.

In this section we briefly exhibit the feature of $\N=2$ abelian gauged supergravity with $B$-field derived from type IIA compactification on the nearly-K\"{a}hler coset space $G_2/SU(3)$.
The details of the derivation can be seen in \cite{Cassani:2008rb, Cassani:2009na}.

\subsection{Profile from coset space $G_2/SU(3)$}
\label{subsect-G2SU3}

First of all let us consider the generic feature of the gauged supergravity 
associated with the type IIA compactification on $G_2/SU(3)$ \cite{KashaniPoor:2007tr, Cassani:2009ck}.
The indices ${\Lambda}$ and $I$ run only $\Lambda = 0,1$ and $I = 0$, respectively.
The following flux charge parameters involve the profile of this compactification:
\begin{gather}
e_{10} \ = \ 2 \sqrt{3 \mathscr{I}}
\, , \ \ \ 
m_{\text{R}}^{\Lambda} e_{\Lambda 0} \ = \ 0
\, , \ \ \ 
m_{\text{R}}^0 \ \neq \ 0
\, , \ \ \ 
e_{\text{R}0} \ \neq \ 0
\, , \label{flux-para-G2SU3}
\end{gather}
whilst other flux charges such as $e_{\Lambda}{}^0$, $e_{00}$, $m_{\text{R}}^1$ and $e_{\text{R}1}$, are zero.
$m_{\text{R}}^0$ is interpreted as the Romans' mass parameter.
The value $\mathscr{I}$ denotes the volume of the coset space.
In this compactification,
the moduli space of the vector multiplet is given by $SU(1,1)/U(1)$.
This is governed by the cubic prepotential ${\cal F}(X)$:
\bsubeq \label{TTT-info-NSR1}
\begin{align}
{\cal F} \ &\equiv \ 
\mathscr{I} \frac{X^1 X^1 X^1}{X^0}
\, .
\end{align}
In terms of the local coordinate $\t \equiv X^1/X^0$, we describe the K\"{a}hler potential $K_{\text{V}}$:
\begin{align}
K_{\text{V}} \ &\equiv \ 
- \log \big[ \i (\ol{X}{}^{\Lambda} {\cal F}_{\Lambda} - X^{\Lambda} \ol{\cal F}_{\Lambda}) \big]
\ = \ 
- \log \big[ - \i \mathscr{I} (\t - \ol{\t})^3 \big]
\, . \label{K-metric}
\end{align}
It is quite useful to introduce the period matrix $\N_{\Lambda \Sigma}$ on this moduli space:
\begin{align}
\N_{\Lambda \Sigma} \ &\equiv \ 
\ol{\cal F}_{\Lambda \Sigma} + 2 \i \frac{(\Im{\cal F})_{\Lambda \Gamma} X^{\Gamma}(\Im{\cal F})_{\Sigma \Delta} X^{\Delta}}{X^{\Pi} (\Im{\cal F})_{\Pi \Xi} X^{\Xi}}
\ = \ 
\frac{\mathscr{I}}{2} \left(
\begin{array}{cc}
\t^2 (\t + 3 \ol{\t}) & - 3 \t (\t + \ol{\t}) \\
- 3 \t (\t + \ol{\t}) & 3 (3 \t + \ol{\t})
\end{array} \right)
\, . \label{TTT-periodN-NSR1}
\end{align}
\esubeq
This pertains to the generalization of the gauge coupling constant and the theta-angle in $\N=2$ supersymmetric system.
In the hyper-sector, there exists only one multiplet.
Notice that the non-vanishing $m_{\text{R}}^0$ makes
the axion field $a$ in the hypermultiplet be dualized to the two-form field $B_{\mu \nu}$, referred to as the $B$-field.
We call the multiplet with the constituents $\{ \varphi, \xi^0, \wt{\xi}_0 , B_{\mu \nu} \}$ the hyper-tensor multiplet.

\subsection{Equations of motion}
\label{EOM-G2SU3}

Here we exhibit the equations of motion for the gauge fields $A_{\mu}^{\Lambda}$, the $B$-field $B_{\mu \nu}$, and the gravitational field $g_{\mu \nu}$, which we will utilize exhaustively in the next section\footnote{See \cite{TK1108} for the Lagrangian and the equations of motion for other bosonic fields.}:
\bsubeq \label{eom-NSR1-G2SU3}
\begin{align}
0 \ &= \ 
\frac{\eps^{\sigma \mu \nu \rho}}{2 \sqrt{-g}}
\del_{\mu} \wt{F}_{\Lambda \nu \rho}
- \frac{\eps^{\sigma \mu \nu \rho}}{2 \sqrt{-g}} \del_{\mu} B_{\nu \rho} 
(e_{\text{R}\Lambda} - e_{\Lambda 0} \xi^0)
- \e^{2 \varphi} e_{\Lambda 0} D^{\sigma} \wt{\xi}_0
\, , \label{eom_A_NSR1-G2SU3} \\
0 \ &= \ 
\frac{1}{\sqrt{-g}} \del_{\mu} \Big( \e^{-4 \varphi} \sqrt{-g} H^{\mu \rho \sigma} \Big)
+ \frac{\eps^{\mu \nu \rho \sigma}}{\sqrt{-g}} \Big[
D_{\mu} \xi^0 D_{\nu} \wt{\xi}_0 - D_{\mu} \wt{\xi}_0 D_{\nu} \xi^0
+ (e_{\text{R}\Lambda} - e_{\Lambda 0} \xi^0) F^{\Lambda}_{\mu \nu}
\Big]
\nn \\
\ & \ \ \ \
+ 2 m_{\text{R}}^{\Lambda} \mu_{\Lambda \Sigma} F^{\Sigma \rho \sigma} 
- \frac{\eps^{\mu \nu \rho \sigma}}{\sqrt{-g}} m_{\text{R}}^{\Lambda} 
\nu_{\Lambda \Sigma} F^{\Sigma}_{\mu \nu} 
\, , \label{eom_B_NSR1-G2SU3} \\
E_{\mu \nu}
\ &\equiv \ 
R_{\mu \nu} - \frac{1}{2} R \, g_{\mu \nu}
\nn \\
\ &= \ 
\frac{1}{4} g_{\mu \nu} \, \mu_{\Lambda \Sigma} F^{\Lambda}_{\rho \sigma} F^{\Sigma \rho \sigma} 
- \mu_{\Lambda \Sigma} F^{\Lambda}_{\mu \rho} F^{\Sigma}_{\nu \sigma} \, g^{\rho \sigma}
- g_{\mu \nu} \, g_{\t \ol{\t}} \del_{\rho} \t \del^{\rho} \ol{\t}
+ 2 g_{\t \ol{\t}} \, \del_{\mu} \t \del_{\nu} \ol{\t}
\nn \\
\ &\ \ \ \ 
- g_{\mu \nu} \, \del_{\rho} \varphi \del^{\rho} \varphi
+ 2 \del_{\mu} \varphi \del_{\nu} \varphi
- \frac{\e^{-4 \varphi}}{24} g_{\mu \nu} \, H_{\rho \sigma \lambda} H^{\rho \sigma \lambda} 
+ \frac{\e^{-4 \varphi}}{4} H_{\mu \rho \sigma} H_{\nu}{}^{\rho \sigma}
\nn \\
\ &\ \ \ \
- \frac{\e^{2 \varphi}}{2} g_{\mu \nu} 
\Big(D_{\rho} \xi^0 D^{\rho} \xi^0 + D_{\rho} \wt{\xi}_0 D^{\rho} \wt{\xi}_0 \Big)
+ \e^{2 \varphi} 
\Big(D_{\mu} \xi^0 D_{\nu} \xi^0 + D_{\mu} \wt{\xi}_0 D_{\nu} \wt{\xi}_0 \Big)
- g_{\mu \nu} V
\, . \label{eom_g_NSR1-G2SU3} 
\end{align}
\esubeq
Here we have introduced various functions:
$\mu_{\Lambda \Sigma} = \Im\N_{\Lambda \Sigma}$ and $\nu_{\Lambda \Sigma} = \Re\N_{\Lambda \Sigma}$ are the generalization of the gauge coupling constant, and the field dependent theta-angle, respectively;
$g_{\t \ol{\t}}$ is the K\"{a}hler metric defined by $g_{\t \ol{\t}} = \del_{\t} \del_{\ol{\t}} K_{\text{V}}$;
and $E_{\mu \nu}$ is the Einstein tensor.
The field strength of the $B$-field is given as $H_{\mu \nu \rho} = 3 \del_{[\mu} B_{\nu \rho]}$.
Notice that, due to the presence of the RR-flux charges $m_{\text{R}}^{\Lambda}$, the gauge field strengths are deformed to \cite{Cassani:2008rb}
\begin{align}
F_{\mu \nu}^{\Lambda} \ = \ 
2 \del_{[\mu} A_{\nu]}^{\Lambda} + m_{\text{R}}^{\Lambda} B_{\mu \nu}
\, . \label{Stuckelberg-type}
\end{align}
The covariant derivatives of the scalar fields $\xi^0$ and $\wt{\xi}_0$ are given by
\begin{align}
D_{\mu} \xi^0 \ &= \ 
\del_{\mu} \xi^0 
\, , \ls
D_{\mu} \wt{\xi}_0 \ = \ 
\del_{\mu} \wt{\xi}_0 - e_{\Lambda 0} A_{\mu}^{\Lambda}
\, . \label{covderiv-NSR1}
\end{align}
They are derived from abelian gauge symmetry of the RR potentials in the ten-dimensional type IIA theory, and from the geometrical structure of the six-dimensional internal space \cite{Cassani:2008rb}.
Due to the absence of the flux charges $e_{\Lambda}{}^0$ on the coset space $G_2/SU(3)$, 
the covariant derivative $D\xi^0$ is reduced to the ordinary derivative 
\cite{Cassani:2009ck}.
The dual gauge field strengths $\wt{F}_{\Lambda \mu \nu}$ are
\begin{gather}
\wt{F}_{\Lambda \mu \nu} \ \equiv \ 
\nu_{\Lambda \Sigma} F^{\Sigma}_{\mu \nu}
+ \frac{\sqrt{-g}}{2} \eps_{\mu \nu \rho \sigma} \, \mu_{\Lambda \Sigma} F^{\Sigma \rho \sigma}
\, ,  \label{dual-F-NSR1}
\end{gather}
where we used the constant tensor whose normalization is $\eps_{0123} = + 1$, and its contravariant form $\eps^{0123} = - 1$ in a generic curved spacetime. 
The scalar potential $V$ is given as \cite{Cassani:2009na}
\bsubeq
\begin{gather}
V \ = \ 
g^{\t \ol{\t}} {\cal D}_{\t} {\cal P}_+ {\cal D}_{\ol{\t}} \ol{\cal P}_+ 
+ g^{\t \ol{\t}} {\cal D}_{\t} {\cal P}_3 {\cal D}_{\ol{\t}} \ol{\cal P}_3 
- 2 |{\cal P}_+|^2 + |{\cal P}_3|^2
\, . \label{V-NSR1-G2SU3} 
\end{gather}
Due to the absence of the flux charges $e_{\Lambda}{}^0$, 
the scalar field $\wt{\xi}_0$ does not contribute to the scalar potential $V$.
The triplet of the Killing prepotentials ${\cal P}_a$ \cite{Cassani:2009na} is explicitly described as
\begin{gather}
{\cal P}_+ \ = \ 
- 2 \e^{\varphi} L^{1} e_{1 0}
\, , \ls
{\cal P}_- \ = \ 
- 2 \e^{\varphi} L^{1} e_{1 0}
\, , \ls
{\cal P}_3 \ = \ 
\e^{2 \varphi} \big[ L^{0} e_{\text{R}0} - L^{1} e_{1 0} \xi^0 - M_{0} m_{\text{R}}^{0} \big]
\, , \label{KillingPrepot} \\
L^0 \ = \ \e^{K_{\text{V}}/2}
\, , \ls
L^1 \ = \ \t \, \e^{K_{\text{V}}/2}
\, , \ls
M_0 \ = \ 
- \mathscr{I} \, \t^3 \e^{K_{\text{V}}/2}
\, , \\
{\cal D}_{\t} {\cal P}_a \ = \ 
\Big( \del_{\t} + \half \del_{\t} K_{\text{V}} \Big) {\cal P}_a
\label{killingprepot}
\end{gather}
\esubeq

\subsection{Static setup}

So far we exhibited generic feature of the gauged supergravity.
Here we prepare the static, spherically symmetric metric:
\bsubeq \label{static}
\begin{gather}
\d s^2 \ = \ 
- \e^{2A(r)} \d t^2 + \e^{-2A(r)} \d r^2 + \e^{2C(r)} r^2 \big( \d \theta^2 + \sin^2 \theta \, \d \phi^2 \big)
\, . \label{metric}
\end{gather}
We impose the time-independent condition on an arbitrary field $X$ such as
\begin{align}
0 \ &= \ 
\del_t X 
\, . \label{static-r5}
\end{align}
\esubeq
The electric and magnetic charges are defined in terms of the field strengths:
\bsubeq \label{EM-charges-r5}
\begin{align}
p^{\Lambda} \ &\equiv \
\frac{1}{4 \pi} \int \d \theta \d \phi \, F^{\Lambda}_{\theta \phi} 
\, , \label{p-r5} \\
q_{\Lambda} \ &\equiv \ 
\frac{1}{4 \pi} \int \d \theta \d \phi \, \wt{F}_{\Lambda \theta \phi} 
\ = \
\frac{1}{4 \pi} \int \d \theta \d \phi \,
\Big( \nu_{\Lambda \Sigma} F^{\Sigma}_{\theta \phi} 
+ \sqrt{-g} \, \mu_{\Lambda \Sigma} F^{\Sigma t r} \Big)
\, . \label{q-r5} 
\end{align}
\esubeq
Since we concentrate on the static configuration of the electric field and the magnetic fields, 
we consider the following set of relations between the gauge fields and the $B$-field: 
\bsubeq \label{fg-r5}
\begin{align}
F^{\Lambda}_{\theta \phi} \ &= \ 
\del_{\theta} A_{\phi}^{\Lambda} - \del_{\phi} A_{\theta}^{\Lambda}
+ m_{\text{R}}^{\Lambda} B_{\theta \phi}
\ \equiv \ 
f^{\Lambda} (\theta, \phi) \sin \theta
\, , \\
F^{\Lambda}_{t r} \ &= \ 
- \del_r A_t^{\Lambda} + m_{\text{R}}^{\Lambda} B_{tr}
\ \equiv \ 
\frac{\e^{-2C}}{r^2} g^{\Lambda} (\theta, \phi) 
\, ,
\end{align}
for the non-vanishing components of the field strengths, and  
\begin{align}
0 \ &= \ 
\del_r F^{\Lambda}_{\theta \phi}
\, , \ls
0 \ = \ 
\del_r \wt{F}_{\Lambda \theta \phi}
\, , \label{r-indept-fg-r5} \\
0 \ &= \ F^{\Lambda}_{\phi r}
\ = \ 
\del_{\phi} A_r^{\Lambda} - \del_r A_{\phi}^{\Lambda} + m_{\text{R}}^{\Lambda} B_{\phi r}
\, , \\
0 \ &= \ F^{\Lambda}_{r \theta}
\ = \ 
\del_r A_{\theta}^{\Lambda} - \del_{\theta} A_r^{\Lambda} + m_{\text{R}}^{\Lambda} B_{r \theta}
\, , \\
0 \ &= \ F^{\Lambda}_{t \theta}
\ = \ 
- \del_{\theta} A_t^{\Lambda} + m_{\text{R}}^{\Lambda} B_{t \theta}
\, , \\
0 \ &= \ F^{\Lambda}_{t \phi}
\ = \ 
- \del_{\phi} A_t^{\Lambda} + m_{\text{R}}^{\Lambda} B_{t \phi}
\, ,
\end{align}
\esubeq
for the vanishing components, respectively.
Note that the fields are independent of time coordinate because we consider the static configuration (\ref{static-r5}).

\section{Analysis}
\label{sect-analysis}

In this section we carefully analyze the equations of motion (\ref{eom-NSR1-G2SU3}) under the static setup (\ref{static}).
Since the $B$-field is incorporated into the equation of motion and the gauge field strengths (\ref{Stuckelberg-type}), 
we can find a series of restrictions on the $B$-field via the equation of motion for the gauge fields.
The equation of motion for the $B$-field further gives rise to differential equations among the set of functions $\{ f^{\Lambda} (\theta, \phi), g^{\Lambda} (\theta, \phi), C(r) \}$, and the scalar fields $\{ \varphi, \xi^0 , \wt{\xi}_0\}$ of the hyper-tensor multiplet.
Utilizing these restrictions, we evaluate the Einstein equation in the static, spherically symmetric (AdS) black hole spacetime.
Recombining the components of the Einstein equation, 
we finally obtain the (covariantly) constant condition.

\subsection{Equation of motion for gauge fields}

First we evaluate the equation of motion for the gauge fields (\ref{eom_A_NSR1-G2SU3}).
Each component provides a powerful constraint among the fields and the functions:
\bsubeq \label{eom_A-r5}
\begin{alignat}{2}
\text{(\ref{eom_A_NSR1-G2SU3})}^{\sigma = t} &: &\ \ \ 
0 \ &= \ 
\frac{\e^{-2C}}{r^2 \sin \theta}
(e_{\text{R}\Lambda} - e_{\Lambda 0} \xi^0) H_{r \theta \phi}
+ e_{\Lambda 0} \, \e^{2 \varphi - 2A} D_t \wt{\xi}_0 
\, , \label{eom_At-r5} \\
\text{(\ref{eom_A_NSR1-G2SU3})}^{\sigma = r} &: &\ \ \ 
0 \ &= \ 
- (e_{\text{R}\Lambda} - e_{\Lambda 0} \xi^0)
\frac{\e^{-2C}}{r^2 \sin \theta} H_{\theta \phi t}
- e_{\Lambda 0} \, \e^{2 \varphi + 2 A} D_r \wt{\xi}_0
\, , \label{eom_Ar-r5} \\
\text{(\ref{eom_A_NSR1-G2SU3})}^{\sigma = \theta} &: &\ \ \ 
0 \ &= \ 
- \frac{\e^{-2C}}{r^2 \sin \theta}
\Big[
\del_{\phi} \wt{F}_{\Lambda tr}
- (e_{\text{R}\Lambda} - e_{\Lambda 0} \xi^0) H_{\phi tr} 
+ e_{\Lambda 0} \, \e^{2 \varphi} D_{\theta} \wt{\xi}_0 \, \sin \theta
\Big]
\, , \label{eom_Ath-r5} \\
\text{(\ref{eom_A_NSR1-G2SU3})}^{\sigma = \phi} &: &\ \ \ 
0 \ &= \ 
\frac{\e^{-2C}}{r^2 \sin \theta}
\Big[
\del_{\theta} \wt{F}_{\Lambda t r}
- (e_{\text{R}\Lambda} - e_{\Lambda 0} \xi^0) H_{\theta tr} 
- e_{\Lambda 0} \, \frac{\e^{2 \varphi}}{\sin \theta} D_{\phi} \wt{\xi}_0
\Big]
\, . \label{eom_Ap-r5}
\end{alignat}
\esubeq
Multiplying $m_{\text{R}}^{\Lambda}$ to the above equations and using the identity $m_{\text{R}}^{\Lambda} e_{\Lambda 0} = 0$ due to the flux compactification \cite{Cassani:2008rb}, 
we extract the following forms:
\bsubeq \label{B1-r5}
\begin{align}
0
\ &= \ 
m_{\text{R}}^{\Lambda} e_{\text{R}\Lambda} \, H_{r \theta \phi}
\, , \\
0 \ &= \ 
m_{\text{R}}^{\Lambda} e_{\text{R}\Lambda} \, H_{\theta \phi t}
\, , \\
m_{\text{R}}^{\Lambda} \del_{\phi} \wt{F}_{\Lambda tr} \ &= \ 
m_{\text{R}}^{\Lambda} e_{\text{R}\Lambda} \, H_{\phi tr}
\, , \\
m_{\text{R}}^{\Lambda} \del_{\theta} \wt{F}_{\Lambda tr} \ &= \ 
m_{\text{R}}^{\Lambda} e_{\text{R}\Lambda} H_{\theta t r}
\, .
\end{align}
\esubeq
Furthermore, the expressions (\ref{fg-r5})
enable us to rewrite the components of the three-form $H_{\mu \nu \rho}$ 
in terms of the functions $f^{\Lambda} (\theta, \phi)$ and $g^{\Lambda} (\theta, \phi)$:
\bsubeq 
\begin{align}
m_{\text{R}}^{\Lambda} e_{\text{R}\Lambda} \, H_{\phi tr}
\ = \ 
m_{\text{R}}^{\Lambda} e_{\text{R}\Lambda} 
\Big[ \del_{\phi} B_{tr} + \del_r B_{\phi t} \Big]
\ &= \ 
e_{\text{R}\Lambda} \Big[
\del_{\phi} \Big( \del_r A_t^{\Lambda} + \frac{\e^{-2C}}{r^2} g^{\Lambda} (\theta, \phi) \Big)
+\del_r \Big( - \del_{\phi} A_t^{\Lambda} \Big)
\Big]
\nn \\
\ &= \
\frac{\e^{-2C}}{r^2} \del_{\phi} \Big[ e_{\text{R}\Lambda} g^{\Lambda} (\theta, \phi) \Big]
\, , \\
m_{\text{R}}^{\Lambda} e_{\text{R}\Lambda} H_{\theta t r}
\ = \ 
m_{\text{R}}^{\Lambda} e_{\text{R}\Lambda} 
\Big[ \del_r B_{\theta t} + \del_{\theta} B_{tr} \Big]
\ &= \ 
e_{\text{R}\Lambda} \Big[
\del_r \Big( - \del_{\theta} A_t^{\Lambda} \Big)
+ \del_{\theta} \Big( \del_r A_t^{\Lambda} + \frac{\e^{-2C}}{r^2} g^{\Lambda} (\theta, \phi) \Big)
\Big]
\nn \\
\ &= \
\frac{\e^{-2C}}{r^2} \del_{\theta} \Big[ e_{\text{R}\Lambda} g^{\Lambda} (\theta, \phi) \Big]
\, .
\end{align}
\esubeq
Substituting them into (\ref{B1-r5}), we obtain the explicit forms: 
\bsubeq \label{B2-r5}
\begin{align}
H_{\phi t r} \ &= \ 
\frac{1}{m_{\text{R}}^{\Sigma} e_{\text{R}\Sigma}}
\frac{\e^{-2C}}{r^2} 
\del_{\phi} \Big[ e_{\text{R}\Lambda} g^{\Lambda} (\theta, \phi) \Big]
\, , \\
H_{\theta tr} \ &= \ 
\frac{1}{m_{\text{R}}^{\Sigma} e_{\text{R}\Sigma}}
\frac{\e^{-2C}}{r^2} 
\del_{\theta} \Big[ e_{\text{R}\Lambda} g^{\Lambda} (\theta, \phi) \Big]
\, , \\
0 \ &= \ 
\del_{\phi} \Big[
(m_{\text{R}}^{\Lambda} \mu_{\Lambda \Sigma}) f^{\Sigma} (\theta, \phi)
+ \big( m_{\text{R}}^{\Lambda} \nu_{\Lambda \Sigma} - e_{\text{R}\Sigma} \big) g^{\Sigma} (\theta, \phi) \Big]
\, , \label{B2p-r5} \\
0 \ &= \ 
\del_{\theta} \Big[
(m_{\text{R}}^{\Lambda} \mu_{\Lambda \Sigma}) f^{\Sigma} (\theta, \phi)
+ \big( m_{\text{R}}^{\Lambda} \nu_{\Lambda \Sigma} - e_{\text{R}\Sigma} \big) g^{\Sigma} (\theta, \phi) \Big]
\, . \label{B2th-r5}
\end{align}
\esubeq
Under the above description, the equations of motion (\ref{eom_A-r5}) are further reduced to
\bsubeq \label{eom_A2-r5}
\begin{align}
0 \ &= \ 
D_t \wt{\xi}_0
\ = \ 
- e_{\Lambda 0} A_t^{\Lambda}
\, , \label{Dtwtxi-r5} \\
0 \ &= \ 
D_r \wt{\xi}_0
\ = \ 
\del_r \wt{\xi}_0 - e_{\Lambda 0} A_r^{\Lambda}
\, , \\
0 \ &= \ 
\frac{\e^{-2C}}{r^2} \del_{\phi} \Big[
\mu_{\Lambda \Sigma} f^{\Sigma} (\theta, \phi)
+ \Big( \nu_{\Lambda \Sigma} - \frac{e_{\text{R}\Lambda} - e_{\Lambda 0} \xi^0}{m_{\text{R}}^{\Gamma} e_{\text{R}\Gamma}} e_{\text{R}\Sigma} \Big) g^{\Sigma} (\theta, \phi)
\Big]
+ e_{\Lambda 0} \, \e^{2 \varphi} \sin \theta \, D_{\theta} \wt{\xi}_0 
\, , \label{eom_Ath2-r5} \\
0 \ &= \ 
\frac{\e^{-2C}}{r^2} \del_{\theta} \Big[
\mu_{\Lambda \Sigma} f^{\Sigma} (\theta, \phi)
+ \Big( \nu_{\Lambda \Sigma} - \frac{e_{\text{R}\Lambda} - e_{\Lambda 0} \xi^0}{m_{\text{R}}^{\Gamma} e_{\text{R}\Gamma}} e_{\text{R}\Sigma} \Big) g^{\Sigma} (\theta, \phi)
\Big]
- e_{\Lambda 0} \frac{\e^{2 \varphi}}{\sin \theta} D_{\phi} \wt{\xi}_0
\, . \label{eom_Ap2-r5}
\end{align}
\esubeq
The equation (\ref{Dtwtxi-r5}) with the flux charge condition (\ref{flux-para-G2SU3}) imposes a strong condition on $g^{\Lambda} (\theta, \phi)$:
\begin{gather}
e_{\Lambda 0} F^{\Lambda}_{tr}
\ = \ 
- \del_r (e_{\Lambda 0} A_t^{\Lambda})
\ = \ 0
\, , \ls
\therefore \ \ \ 
e_{\Lambda 0} g^{\Lambda} (\theta, \phi) \ = \ 0
\, . \label{eg0-r5} 
\end{gather}
Notice that $g^0 (\theta, \phi)$ still remains non-trivial.

\subsection{Equation of motion for $B$-field}

Next task is to investigate the equation of motion for the $B$-field (\ref{eom_B_NSR1-G2SU3}). 
Owing to the expressions (\ref{fg-r5}) and (\ref{B2-r5}), 
each component of the equation is described in terms of the functions $f^{\Lambda} (\theta, \phi)$ and $g^{\Lambda} (\theta, \phi)$ in the following way:
\bsubeq \label{eom_B-r5}
\begin{alignat}{2}
\text{(\ref{eom_B_NSR1-G2SU3})}^{[\rho \sigma] = [tr]}&: &\ \ \
0 \ &= \ 
- \frac{1}{m_{\text{R}}^{\Sigma} e_{\text{R}\Sigma}}
\frac{\e^{-2C}}{r^2}
\del_{\theta} \Big[
\e^{-4 \varphi} \sin \theta \,
\del_{\theta} \Big( e_{\text{R}\Lambda} g^{\Lambda} (\theta, \phi) \Big)
\Big]
\nn \\
\ &&& \ \ \ \ 
- \frac{1}{m_{\text{R}}^{\Sigma} e_{\text{R}\Sigma}}
\frac{\e^{-2C}}{r^2 \sin \theta}
\del_{\phi} \Big[
\e^{-4 \varphi} \del_{\phi} \Big( e_{\text{R}\Lambda} g^{\Lambda} (\theta, \phi) \Big)
\Big]
\nn \\
\ &&& \ \ \ \ 
+ 2 \Big[
\Big( m_{\text{R}}^{\Lambda} \nu_{\Lambda \Sigma} 
- (e_{\text{R}\Sigma} - e_{\Sigma 0} \, \xi^0) \Big) f^{\Sigma} (\theta , \phi)
- (m_{\text{R}}^{\Lambda} \mu_{\Lambda \Sigma}) g^{\Sigma} (\theta , \phi)
\Big] \sin \theta
\nn \\
\ &&& \ \ \ \ 
- 2 \Big( \del_{\theta} \xi^0 D_{\phi} \wt{\xi}_0 
- D_{\theta} \wt{\xi}_0 \del_{\phi} \xi^0 \Big)
\, , \label{eom_Btr2-r5} \\
\text{(\ref{eom_B_NSR1-G2SU3})}^{[\rho \sigma] = [t\theta]}&: &\ \ \
0 \ &= \ 
\frac{\sin \theta}{m_{\text{R}}^{\Sigma} e_{\text{R}\Sigma}} 
\del_{\theta} \Big( e_{\text{R}\Lambda} g^{\Lambda} (\theta, \phi) \Big)
\del_r \Big( \frac{\e^{-4 \varphi-2C}}{r^2} \Big)
+ 2 \del_r \xi^0 D_{\phi} \wt{\xi}_0
\, , \label{eom_Btth2-r5} \\
\text{(\ref{eom_B_NSR1-G2SU3})}^{[\rho \sigma] = [t\phi]}&: &\ \ \
0 \ &= \ 
\frac{1}{m_{\text{R}}^{\Sigma} e_{\text{R}\Sigma}} 
\frac{1}{\sin \theta} 
\del_{\phi} \Big( e_{\text{R}\Lambda} g^{\Lambda} (\theta, \phi) \Big)
\del_r \Big( \frac{\e^{-4\varphi-2C}}{r^2} \Big)
- 2 \del_r \xi^0 D_{\theta} \wt{\xi}_0
\, , \label{eom_Btp2-r5} \\
\text{(\ref{eom_B_NSR1-G2SU3})}^{[\rho \sigma] = [\theta\phi]}&: &\ \ \
0 \ &= \ 
\frac{2 \e^{-4C}}{r^4 \sin \theta} \Big[
(m_{\text{R}}^{\Lambda} \mu_{\Lambda \Sigma}) f^{\Sigma} (\theta, \phi)
+ (m_{\text{R}}^{\Lambda} \nu_{\Lambda \Sigma} - e_{\text{R}\Sigma}) g^{\Sigma} (\theta , \phi)
\Big]
\, , \label{eom_Bthp-r5} 
\end{alignat}
\esubeq
where we used (\ref{eg0-r5}). 
Then the equations (\ref{B2p-r5}) and (\ref{B2th-r5}) become trivial:
\begin{align}
0 \ &= \ 
(m_{\text{R}}^{\Lambda} \mu_{\Lambda \Sigma}) f^{\Sigma} (\theta, \phi)
+ (m_{\text{R}}^{\Lambda} \nu_{\Lambda \Sigma} - e_{\text{R}\Sigma}) g^{\Sigma} (\theta , \phi)
\, . \label{const-fg2-r5}
\end{align}
We can learn that the covariant derivatives of the RR-axion fields are related to the derivatives of the function $g^{\Lambda} (\theta, \phi)$.
Indeed the equations (\ref{eom_Btth2-r5}) and (\ref{eom_Btp2-r5}) will contribute to the evaluation of the Einstein equation in a crucial way.

\subsection{Einstein equation}

So far we analyzed the equations of motion in terms of arbitrary functions $A(r)$ and $C(r)$ in the static metric (\ref{metric-r5}).
From now on, we focus only on the metric of Reissner-Nordstr\"{o}m-AdS type:
\begin{gather}
\e^{2 A(r)} \ \equiv \ 
\kappa - \frac{2 \eta}{r} + \frac{{\cal Z}^2}{r^2} + \frac{r^2}{\l^2}
\, , \ls
\e^{2 C(r)} \ \equiv \ 1 
\, , \label{metric-r5}
\end{gather}
where the black hole parameters of mass and charges are 
given by $\eta$ and ${\cal Z}^2 = Q_e^2 + Q_m^2$, respectively.
The parameter $\l$ gives 
the negative cosmological constant $\Lambda_{\text{c.c.}} = - 3/\l^2$.

The diagonal components of the Einstein tensor $E_{\mu \nu}$ under the metric (\ref{metric-r5}) become simple:
\bsubeq \label{Einstein-r5}
\begin{align}
g^{tt} E_{tt}
\ &= \ 
\e^{2 A} \Big[
\frac{1}{r^2} (1 - \e^{-2(A+C)}) 
+ \frac{2}{r} (A'+ 3 {C'}) 
+ {C'} (2 A' + 3C') 
+ 2 {C''}
\Big]
\ = \ 
- \frac{{\cal Z}^2}{r^4}
+ \frac{3}{\l^2}
\, , \\
g^{rr} E_{rr}
\ &= \ 
\e^{2 A} \Big[
\frac{1}{r^2} (1 - \e^{-2(A+C)}) + \frac{2}{r}(A' + {C'}) + {C'}(2A' + C') 
\Big]
\ = \ 
- \frac{{\cal Z}^2}{r^4}
+ \frac{3}{\l^2}
\, , \\
g^{\theta \theta} E_{\theta \theta}
\ &= \ 
\e^{2 A} \Big[
\frac{2}{r} (A' + {C'}) + 2 (A')^2 + {C'}(2A' + C') + A'' + {C''} 
\Big]
\ = \ 
\frac{{\cal Z}^2}{r^4}
+ \frac{3}{\l^2}
\, , \\
g^{\phi \phi} E_{\phi \phi}
\ &= \
\e^{2 A} \Big[
\frac{2}{r} (A' + {C'}) + 2 (A')^2 + {C'}(2A' + C') + A'' + {C''} 
\Big] 
\ = \ 
\frac{{\cal Z}^2}{r^4}
+ \frac{3}{\l^2}
\, .
\end{align}
\esubeq
On the other hand, substituting the vanishing conditions (\ref{fg-r5}) and (\ref{B1-r5}),
we see the reduced components of the right-hand side of the Einstein equation (\ref{eom_g_NSR1-G2SU3}):
\bsubeq \label{eom_g-r5}
\begin{align}
g^{tt} E_{tt} \ &= \ 
- \frac{1}{2} \mu_{\Lambda \Sigma} F^{\Lambda}_{tr} F^{\Sigma tr}
+ \frac{1}{2} \mu_{\Lambda \Sigma} F^{\Lambda}_{\theta \phi} F^{\Sigma \theta \phi}
- g_{\t \ol{\t}} \Big( \del_{r} \t \del^{r} \ol{\t}
+ \del_{\theta} \t \del^{\theta} \ol{\t}
+ \del_{\phi} \t \del^{\phi} \ol{\t} \Big)
\nn \\
\ & \ \ \ \ 
- \Big( \del_{r} \varphi \del^{r} \varphi
+ \del_{\theta} \varphi \del^{\theta} \varphi
+ \del_{\phi} \varphi \del^{\phi} \varphi \Big)
+ \frac{\e^{-4 \varphi}}{4} H_{tr \theta} H^{tr \theta}
+ \frac{\e^{-4 \varphi}}{4} H_{tr \phi} H^{tr \phi}
\nn \\
\ & \ \ \ \ 
- \frac{\e^{2 \varphi}}{2} \Big(
\del_r \xi^0 \del^r \xi^0
+ \del_{\theta} \xi^0 \del^{\theta} \xi^0
+ \del_{\phi} \xi^0 \del^{\phi} \xi^0
\Big)
- \frac{\e^{2 \varphi}}{2} \Big(
D_{\theta} \wt{\xi}_0 D^{\theta} \wt{\xi}_0 
+ D_{\phi} \wt{\xi}_0 D^{\phi} \wt{\xi}_0 
\Big)
- V
\, , \label{eom_gt-r5} \\
g^{rr} E_{rr} \ &= \ 
- \frac{1}{2} \mu_{\Lambda \Sigma} F^{\Lambda}_{tr} F^{\Sigma tr}
+ \frac{1}{2} \mu_{\Lambda \Sigma} F^{\Lambda}_{\theta \phi} F^{\Sigma \theta \phi}
- g_{\t \ol{\t}} \Big(
- \del_{r} \t \del^{r} \ol{\t}
+ \del_{\theta} \t \del^{\theta} \ol{\t}
+ \del_{\phi} \t \del^{\phi} \ol{\t}
\Big)
\nn \\
\ & \ \ \ \ 
- \Big( - \del_{r} \varphi \del^{r} \varphi
+ \del_{\theta} \varphi \del^{\theta} \varphi
+ \del_{\phi} \varphi \del^{\phi} \varphi
\Big)
+ \frac{\e^{-4 \varphi}}{4} H_{t r \theta} H^{tr \theta}
+ \frac{\e^{-4 \varphi}}{4} H_{t r \phi} H^{tr \phi}
\nn \\
\ & \ \ \ \ 
- \frac{\e^{2 \varphi}}{2} \Big(
- \del_r \xi^0 \del^r \xi^0
+ \del_{\theta} \xi^0 \del^{\theta} \xi^0
+ \del_{\phi} \xi^0 \del^{\phi} \xi^0
\Big)
- \frac{\e^{2 \varphi}}{2} \Big(
D_{\theta} \wt{\xi}_0 D^{\theta} \wt{\xi}_0 
+ D_{\phi} \wt{\xi}_0 D^{\phi} \wt{\xi}_0 
\Big)
- V
\, , \label{eom_gr-r5} \\
g^{\theta \theta} E_{\theta \theta} 
\ &= \ 
\frac{1}{2} \mu_{\Lambda \Sigma} F^{\Lambda}_{tr} F^{\Sigma tr}
- \frac{1}{2} \mu_{\Lambda \Sigma} F^{\Lambda}_{\theta \phi} F^{\Sigma \theta \phi}
- g_{\t \ol{\t}} \Big(
\del_{r} \t \del^{r} \ol{\t}
- \del_{\theta} \t \del^{\theta} \ol{\t}
+ \del_{\phi} \t \del^{\phi} \ol{\t}
\Big)
\nn \\
\ & \ \ \ \ 
- \Big(
\del_{r} \varphi \del^{r} \varphi
- \del_{\theta} \varphi \del^{\theta} \varphi
+ \del_{\phi} \varphi \del^{\phi} \varphi
\Big)
+ \frac{\e^{-4 \varphi}}{4} H_{tr \theta} H^{tr \theta}
- \frac{\e^{-4 \varphi}}{4} H_{tr \phi} H^{tr \phi}
\nn \\
\ & \ \ \ \ 
- \frac{\e^{2 \varphi}}{2} \Big(
\del_r \xi^0 \del^r \xi^0
- \del_{\theta} \xi^0 \del^{\theta} \xi^0
+ \del_{\phi} \xi^0 \del^{\phi} \xi^0
\Big)
- \frac{\e^{2 \varphi}}{2} \Big(
- D_{\theta} \wt{\xi}_0 D^{\theta} \wt{\xi}_0 
+ D_{\phi} \wt{\xi}_0 D^{\phi} \wt{\xi}_0 
\Big)
- V
\, , \label{eom_gth-r5} \\
g^{\phi \phi} E_{\phi \phi} 
\ &= \ 
\frac{1}{2} \mu_{\Lambda \Sigma} F^{\Lambda}_{tr} F^{\Sigma tr}
- \frac{1}{2} \mu_{\Lambda \Sigma} F^{\Lambda}_{\theta \phi} F^{\Sigma \theta \phi}
- g_{\t \ol{\t}} \Big(
\del_{r} \t \del^{r} \ol{\t}
+ \del_{\theta} \t \del^{\theta} \ol{\t}
- \del_{\phi} \t \del^{\phi} \ol{\t}
\Big)
\nn \\
\ & \ \ \ \ 
- \Big( 
\del_{r} \varphi \del^{r} \varphi
+ \del_{\theta} \varphi \del^{\theta} \varphi
- \del_{\phi} \varphi \del^{\phi} \varphi
\Big)
- \frac{\e^{-4 \varphi}}{4} H_{tr \theta} H^{tr \theta}
+ \frac{\e^{-4 \varphi}}{4} H_{tr \phi} H^{tr \phi}
\nn \\
\ & \ \ \ \ 
- \frac{\e^{2 \varphi}}{2} \Big(
\del_r \xi^0 \del^r \xi^0
+ \del_{\theta} \xi^0 \del^{\theta} \xi^0
- \del_{\phi} \xi^0 \del^{\phi} \xi^0
\Big) 
- \frac{\e^{2 \varphi}}{2} \Big(
D_{\theta} \wt{\xi}_0 D^{\theta} \wt{\xi}_0 
- D_{\phi} \wt{\xi}_0 D^{\phi} \wt{\xi}_0 
\Big)
- V
\, . \label{eom_gp-r5} 
\end{align}
\esubeq
Combining (\ref{Einstein-r5}) with (\ref{eom_g-r5}), 
we evaluate the derivatives of the fields.
First, the difference between the time component and the radial component gives
\begin{align}
g^{tt} E_{tt} - g^{rr} E_{rr}
\ = \ 
0 \ &= \ 
-2 \e^{2A(r)} \Big[ g_{\t \ol{\t}} |\del_r \t|^2
+ (\del_r \varphi)^2
+ \frac{\e^{2 \varphi}}{2} (\del_r \xi^0)^2
\Big]
\, . \label{gt-gr-r5}
\end{align}
This is a strong condition. 
Because each term in the right-hand side is positive semi-definite, 
we immediately obtain
\begin{align}
0 \ &= \ \del_r \t
\, , \ls
0 \ = \ \del_r \varphi
\, , \ls 
0 \ = \ \del_r \xi^0
\, , \label{t-d-xi-r5}
\end{align}
This makes the derivative of the function $g^{\Lambda} (\theta, \phi)$ vanish through the equations (\ref{eom_Btth2-r5}) and (\ref{eom_Btp2-r5}): 
\begin{gather*}
0 \ = \ \del_{\theta} \Big( e_{\text{R}\Lambda} g^{\Lambda} (\theta, \phi) \Big)
\, , \ls
0 \ = \ \del_{\phi} \Big( e_{\text{R}\Lambda} g^{\Lambda} (\theta, \phi) \Big)
\, , 
\end{gather*}
which implies that $g^{\Lambda}$ is a constant and the components of the three-form $H$ in (\ref{B2-r5}) vanish:
\begin{align}
g^{\Lambda} \ = \ \text{(constant)}
\, , \ls
0 \ = \ H_{tr \theta} \ = \ H_{tr \phi}
\, . \label{H0-r5} 
\end{align}
Combining (\ref{B1-r5}), 
we find that all the components of $H_{\mu \nu \rho}$ vanish once the static condition is imposed on the system.
In a similar way, we also evaluate the following three equations: 
\bsubeq \label{gr-gth-gp-r5}
\begin{align}
g^{rr} E_{rr} + g^{\theta \theta} E_{\theta \theta} \ = \ 
\frac{6}{\l^2}
\ &= \ 
- \frac{2}{r^2 \sin^2 \theta}
\Big[ g_{\t \ol{\t}} |\del_{\phi} \t|^2
+ (\del_{\phi} \varphi)^2
+ \frac{\e^{2 \varphi}}{2} (\del_{\phi} \xi^0)^2
+ \frac{\e^{2 \varphi}}{2} (D_{\phi} \wt{\xi}_0)^2
\Big]
- 2 V
\, , \label{gr+gth-r5} \\
g^{rr} E_{rr} - g^{\theta \theta} E_{\theta \theta} \ = \ 
- \frac{2 {\cal Z}^2}{r^4}
\ &= \ 
\frac{1}{r^4} \mu_{\Lambda \Sigma} \Big[
f^{\Lambda} (\theta, \phi) f^{\Sigma} (\theta, \phi)
+ g^{\Lambda} g^{\Sigma} \Big]
\nn \\
\ & \ \ \ \ 
- \frac{2}{r^2} \Big[
g_{\t \ol{\t}} |\del_{\theta} \t|^2
+ (\del_{\theta} \varphi)^2
+ \frac{\e^{2 \varphi}}{2} (\del_{\theta} \xi^0)^2
+ \frac{\e^{2 \varphi}}{2} (D_{\theta} \wt{\xi}_0)^2
\Big]
\, , \label{gr-gth-r5} \\
g^{\theta \theta} E_{\theta \theta} - g^{\phi \phi} E_{\phi \phi}
\ = \ 
0 \ &= \ 
\frac{1}{r^2}
\Big[
g_{\t \ol{\t}} |\del_{\theta} \t|^2
+ (\del_{\theta} \varphi)^2
+ \frac{\e^{2 \varphi}}{2} (\del_{\theta} \xi^0)^2
+ \frac{\e^{2 \varphi}}{2} (D_{\theta} \wt{\xi}_0)^2
\Big]
\nn \\
\ & \ \ \ \ 
- \frac{1}{r^2 \sin^2 \theta}
\Big[
g_{\t \ol{\t}} |\del_{\phi} \t|^2
+ (\del_{\phi} \varphi)^2
+ \frac{\e^{2 \varphi}}{2} (\del_{\phi} \xi^0)^2
+ \frac{\e^{2 \varphi}}{2} (D_{\phi} \wt{\xi}_0)^2
\Big]
\, . \label{gth-gp-r5} 
\end{align}
\esubeq
Note that we have used the vanishing condition of all the components of $H_{\mu \nu \rho}$.
Since the scalar fields $\t$, $\varphi$ and $\xi^0$ are independent of the radial coordinate $r$ (\ref{t-d-xi-r5}), the period matrix $\N_{\Lambda \Sigma} = \nu_{\Lambda \Sigma} + \i \mu_{\Lambda \Sigma}$ (\ref{TTT-periodN-NSR1}) are also independent of $r$. 
The scalar potential $V$ defined in (\ref{V-NSR1-G2SU3}) does not depend on $r$, either.
Then the square bracket in the right-hand side of (\ref{gr+gth-r5})
and the second line in the right-hand side of (\ref{gr-gth-r5}) must vanish:
\begin{align}
0 \ &= \ 
g_{\t \ol{\t}} |\del_{\theta} \t|^2 
+ (\del_{\theta} \varphi)^2
+ \frac{\e^{2 \varphi}}{2} (\del_{\theta} \xi^0)^2
+ \frac{\e^{2 \varphi}}{2} (D_{\theta} \wt{\xi}_0)^2
\nn \\
\ &= \ 
g_{\t \ol{\t}} |\del_{\phi} \t|^2 
+ (\del_{\phi} \varphi)^2
+ \frac{\e^{2 \varphi}}{2} (\del_{\phi} \xi^0)^2
+ \frac{\e^{2 \varphi}}{2} (D_{\phi} \wt{\xi}_0)^2
\, . 
\end{align}
This condition satisfies (\ref{gth-gp-r5}) consistently.
Because each term in this equation is positive semi-definite,
we obtain 
\bsubeq \label{t-d-xi-Dwtxi-r5}
\begin{alignat}{4}
0 \ &= \ \del_{\theta} \t \ &&= \ \del_{\phi} \t
\, , &\ls
0 \ &= \ \del_{\theta} \varphi \ &&= \ \del_{\phi} \varphi
\, , \\
0 \ &= \ \del_{\theta} \xi^0 \ &&= \ \del_{\phi} \xi^0
\, , &\ls
0 \ &= \ D_{\theta} \wt{\xi}_0 \ &&= \ D_{\phi} \wt{\xi}_0
\, .
\end{alignat}
\esubeq

Due to the constant condition of $g^{\Lambda}$ and $\t$, the equation (\ref{const-fg2-r5}) tells us that $f^{\Lambda}$ becomes constant.
Their values are described in terms of the electric and magnetic charges through (\ref{EM-charges-r5}):
\begin{align}
f^{\Lambda} \ &= \ p^{\Lambda}
\, , \ls
g^{\Lambda} \ = \ 
- (\mu^{-1})^{\Lambda \Sigma} \big( q_{\Sigma} - \nu_{\Sigma \Gamma} \, p^{\Gamma} \big)
\, . \label{fg-final-r5}
\end{align} 

Let us summarize the analysis.
Assuming only the static configuration (\ref{static-r5}), (\ref{static-r5}), and (\ref{fg-r5}),
we obtain the (covariantly) constant forms of the scalar fields and 
the constant gauge field strengths as showed in (\ref{B1-r5}), (\ref{eom_A2-r5}), (\ref{t-d-xi-r5}), (\ref{H0-r5}), (\ref{t-d-xi-Dwtxi-r5}) and (\ref{fg-final-r5}):
\bsubeq  \label{CCC-NSR1}
\begin{gather}
0 \ = \ H_{\mu \nu \rho} \ = \ 3 \del_{[\mu} B_{\nu \rho]}
\, , \ls
0 \ = \ D_{\mu} \wt{\xi}_0
\, , \ls
0 \ = \ D_{\mu} \xi^0 \ = \ \del_{\mu} \xi^0
\, , \ls
0 \ = \ \del_{\mu} \t
\, , \\
F^{\Lambda}_{\theta \phi} \ = \ p^{\Lambda} \sin \theta
\, , \ls
F^{\Lambda}_{tr} \ = \ 
- \frac{1}{r^2} (\mu^{-1})^{\Lambda \Sigma} \big( q_{\Sigma} - \nu_{\Sigma \Gamma} \, p^{\Gamma} \big)
\, .
\end{gather}
\esubeq
This leads to the black hole parameters in such a way as
\bsubeq \label{BH-parameters}
\begin{align}
{\cal Z}^2 \ &= \ 
- \frac{\mu_{\Lambda \Sigma}}{2} \Big[
f^{\Lambda} f^{\Sigma} + g^{\Lambda} g^{\Sigma}
\Big]
\nn \\
\ &= \ 
- \half \Big[
p^{\Lambda} \mu_{\Lambda \Sigma} \, p^{\Sigma}
+ \big( q_{\Lambda} - \nu_{\Lambda \Gamma} \, p^{\Gamma} \big) 
(\mu^{-1})^{\Lambda \Sigma} 
\big( q_{\Sigma} - \nu_{\Sigma \Delta} \, p^{\Delta} \big)
\Big]
\, , \\
\Lambda_{\text{c.c.}} \ &= \ 
- \frac{3}{\l^2}
\ = \ 
V
\, .
\end{align}
\esubeq

\subsection{Constant solution}

Let us proceed the analysis under the field configuration (\ref{CCC-NSR1})\footnote{The essential points have already been discussed in \cite{TK1108}.}.
The covariantly constant condition $0 = D_{\mu} \wt{\xi}_0$ in (\ref{CCC-NSR1}) gives the vanishing condition of the field strengths:
\begin{align}
0 \ &= \ [ \del_{\mu} , \del_{\nu} ] \wt{\xi}_0
\ = \ e_{\Lambda 0} F^{\Lambda}_{\mu \nu}
\, ,
\end{align}
i.e., $F^1_{\mu \nu} = 0$. 
Here we again used the flux charge condition $m_{\text{R}}^{\Lambda} e_{\Lambda 0} = 0$ in (\ref{flux-para-G2SU3}). 
The above condition implies
\begin{align}
0 \ &= \ p^1
\, , \ls
0 \ = \ 
(\mu^{-1})^{10} q_0 + (\mu^{-1})^{11} q_1 - (\mu^{-1} \nu)^1{}_0 \, p^0
\, . \label{pq1-r5}
\end{align}
Substituting this into (\ref{eom_Btr2-r5}), we obtain 
\bsubeq \label{pq2-r5}
\begin{align}
p^0 \ &= \ \frac{m_{\text{R}}^0}{e_{\text{R}0}} q_0
\, , \\
(\mu^{-1})^{11} q_1 \ &= \ 
- q_0 \Big[
(\mu^{-1})^{10} - \Big( \frac{m_{\text{R}}^0}{e_{\text{R}0}} \Big) (\mu^{-1} \nu)^1{}_0
\Big]
\, , \\
0 \ &= \ 
q_0 \Big[
\Big( (\mu^{-1})^{11} (\mu^{-1})^{00} - [(\mu^{-1})^{01}]^2 \Big)
- 2 \Big( \frac{m_{\text{R}}^0}{e_{\text{R}0}} \Big)
\Big( (\mu^{-1})^{11} (\mu^{-1} \nu)^0{}_0 - (\mu^{-1})^{01} (\mu^{-1} \nu)^1{}_0 \Big)
\nn \\
\ &\LS
+ \Big( \frac{m_{\text{R}}^0}{e_{\text{R}0}} \Big)^2
\Big( (\mu^{-1})^{11} (\mu + \nu \mu^{-1} \nu)_{00} 
- [(\mu^{-1} \nu)^1{}_0 ]^2 \Big)
\Big]
\, . \label{pq2-3-r5}
\end{align}
\esubeq
Since the value in the square bracket of (\ref{pq2-3-r5}) is non-zero \cite{TK1108}, the electric charge $q_0$ must vanish.
Substituting this into the above equations, we eventually find that all the charges must be zero:
\begin{align}
q_0 \ &= \ 0
\, , \ls
q_1 \ = \ 0
\, , \ls
p^0 \ = \ 0
\, , \ls
p^1 \ = \ 0
\, .
\end{align}
The gauge field strengths given in (\ref{CCC-NSR1}) also vanish.
This leads to the vanishing black hole charges ${\cal Z}^2 = 0$.
Then we conclude that the static, spherically symmetric configuration in the gauged supergravity derived from type IIA theory compactified on $G_2/SU(3)$ provides only neutral solutions such as AdS vacua or Schwarzschild-AdS black holes \cite{TK1108}.


\section{Conclusion}
\label{sect-conclusion}

In this work we studied static, spherically symmetric, asymptotically AdS 
black hole solutions in four-dimensional $\N=2$ gauged supergravity in the presence of one vector multiplet and one hyper-tensor multiplet.
This system is associated with the type IIA theory compactified on the nearly-K\"{a}hler coset space $G_2/SU(3)$.
The Romans' mass yields the St\"{u}ckelberg-type deformation in the gauge field strengths. 
Then we found an intrinsic relation between the gauge fields $A_{\mu}^{\Lambda}$ and the $B$-field.
Eventually all the scalar fields should satisfy the (covariantly) constant condition, and the $B$-field must be closed.
Furthermore, the (covariantly) constant condition leads to the vanishing black hole charges. 
It turns out that only the possible solutions are AdS vacua or Schwarzschild-AdS black hole as analyzed in \cite{TK1108}.

In the main analysis
we fixed the sign of the cosmological constant to be negative in (\ref{metric-r5}).
But this did not affect the evaluation of the scalar fields in the Einstein equation (\ref{gr+gth-r5}).
In addition, the topology of the horizon is not crucial in the equations (\ref{gr-gth-r5}) and (\ref{gth-gp-r5}). 
The primary reason was that the $H_{\mu \nu \rho}$ did vanish through the equations (\ref{eom_B-r5}) and (\ref{gt-gr-r5}).
This eventually removed the dependence of the $B$-field in the equations (\ref{gr-gth-gp-r5}), and the independence of the angular coordinates (\ref{t-d-xi-Dwtxi-r5}) were realized.
Thus the result (\ref{CCC-NSR1}) could be applied to static, asymptotically flat (or de Sitter) black holes with unusual topology such as two-torus or hyperbolic surface (see, for instance, \cite{Caldarelli:1998hg}).

This work reveals that,
as far as we concern the metric (\ref{metric}) with (\ref{metric-r5}),
time-independent configurations must be forbidden
to build a charged (AdS) black hole in the presence of the hyper-tensor multiplet with the St\"{u}ckelberg-type deformation.
It seems to be inevitable to search time-dependent configurations.
A typical one is the stationary, rotating charged black hole referred to as the Kerr-Newman-(AdS) black hole.


}
\end{document}